\documentclass[conf]{new-aiaa}

\usepackage{graphicx}
\usepackage{amsmath}
\usepackage[version=4]{mhchem}
\usepackage{siunitx}
\usepackage{longtable,tabularx}
\usepackage{lipsum}
\usepackage{graphicx}
\usepackage[caption=false]{subfig}
\usepackage{float}
\usepackage{stmaryrd}
\usepackage{varioref}
\usepackage{tikz}
\usetikzlibrary{shapes.geometric,arrows}
\usepackage{booktabs}
\usepackage{gensymb}
\usepackage{graphicx}
\usepackage{amssymb}
\usepackage[dvips]{epsfig}
\usepackage{epsfig} 
\usepackage{epstopdf}
\usepackage{latexsym}
\usepackage{amsmath} 
\usepackage{amstext}
\usepackage{amsfonts}
\usepackage{amsopn}
\usepackage{fixltx2e}
\usepackage{pifont}
\usepackage{sidecap,wrapfig,psfrag,array,tabularx}
\usepackage{amsfonts,amsopn,amstext,amsfonts,amsbsy}
\usepackage{setspace}
\usepackage{mathtools}

\newcommand{\bi}{\begin{itemize}}
	\newcommand{\ei}{\end{itemize}}
\title{Modeling and Simulation of UAV Carrier Landings}

\author{Gaurav Misra\footnote{Ph.D. Candidate, Mechanical and Aerospace Engineering and AIAA Student Member.}, Tianyu Gao\footnote{Ph.D. Student, Mechanical and Aerospace Engineering}, and Xiaoli Bai\footnote{Assistant Professor, Mechanical and Aerospace Engineering and AIAA Senior Member.}                    }
\affil{Rutgers, The State University of New Jersey, Piscataway, NJ, 08854}
\begin{document}

\maketitle

\begin{abstract}
With UAVs’ promising capabilities to increase operation flexibility and reduce mission cost, we are exploiting the automated carrier-landing performance advancement that can be achieved by fixed-wing UAVs. To demonstrate such potentials, in this paper, we investigate two key metrics, namely, flight path control performance, and reduced approach speeds for UAVs based on the F/A-18 High Angle of Attack (HARV) model. The landing control architecture consists of an auto-throttle, a stability augmentation system, glideslope and approach track controllers. The performance of the control model is  tested using Monte Carlo simulations under a range of environmental uncertainties including atmospheric turbulence consisting of wind shear, discrete and continuous wind gusts, and carrier airwakes. Realistic deck motion is considered where the standard deck motion time histories under the Systematic Characterization of the Naval Environment (SCONE) program released by the Office of Naval Research (ONR) are used. We numerically demonstrate the limiting approach conditions which allow for successful carrier landings and factors affecting it's performance. 
\end{abstract}

\section{Introduction}
The highly demanding task of landing a high-performance aircraft on a carrier has been significantly researched and developed since January 1911 when Eugene Ely landed a biplane aboard on the USS Pennsylvania. Shipboard landing requires an aircraft to land on a pitching and rolling deck in highly turbulent ship airwakes; the landing area is very small and the landing needs to be so precise that the landing error must remain within one foot. Moreover, the landing often has to be performed at night and in inclement weather.  

Although automatic take-off and landing technology has been tested using piloted aircraft such as F/A-18E/F~\cite{durand_analysis_1964, johnson2001test}, the full potential of emerging unmanned air vehicles (UAVs) has not yet been systematically explored and thoroughly investigated for aircraft automated carrier landing. For example, although a low approach speed is highly desired for reasons such as to reduce the loads imposed on the arresting wires and on the aircraft, dependent on the existing flight control system, the current approach speed is required not be less than 110 \% of the stall~\cite{rudowsky2002review}. Although this stall margin criterion has been reported to be inadequate and difficult to justify, we have not found a rigorous study on the possible minimum approach speed. In addition, atmospheric and carrier induced turbulence directly impact the approach conditions. Therefore, reduced approach speeds under turbulence needs further investigation.

Eliminating the factor of pilots from the flight control system design avoids many inherent difficulties for manned aircraft because of crews’ operational and physical constraints and introduces a wide range of otherwise-non-existing flexibilities and potential advantages to be exploited for optimizing the carrier landing processes. Together with the advantage of using many highly mature technologies gained over decades of manned aircraft development, UAVs are expected to achieve performance levels significantly beyond what piloted aircraft could possibly accomplish. 
We are currently exploiting the landing performance advancement that can be achieved by fixed-wing UAVs. The potential benefits include: reduced approach speed closer to stall, reduced sink rate approach near the ship, reduced sink rate at touchdown, reduction of the landing position deviation from the arresting wire, and reduction of the flight path deviation from the reference.  To demonstrate such potentials, we develop baseline aircraft models with baseline flight controls representative of the F/A-18 High Angle of Attack (HARV) model, which will be used to compare the carrier landing performance between the current technology and the advanced concepts proposed in this research. 

Although recent literature on automated carrier landing looks at advanced control techniques such as $\ell_{1}$ adaptive control~\cite{ramesh2016autonomous}, disturbance rejection control~\cite{wang_adaptive_2016}, preview control~\cite{zhen_preview_2018}, and stochastic model predictive control~\cite{misra2018stochastic}, in this paper, the focus is on numerical investigation of flight performance and reduced approach speeds under baseline proportional-integral-derivative (PID) feedback control laws. This approach is taken since current operational control architectures are largely PID based. In addition, in largely all of the current available literature, the usual assumption is a fixed approach condition, with an approach speed typically in the range of $220-250$ ft/s and a fixed descent glideslope of $2.5-4$ $\deg$. The main contributions of this paper are the rigorous numerical verification of the flight control architecture under a range of environmental conditions which include low intensity atmospheric turbulence, carrier airwakes and deck motion. In addition, we numerically demonstrate the limiting approach speed at which carrier landing can be conducted under the same environmental setup. The landing performance is assessed by studying the deck landing dispersion, final altitude error, final glideslope, and lateral state errors.

The paper outline is as follows. Section~\ref{simulation_models} focuses on the baseline simulation including the aircraft model, the control laws consisting of a stability augmentation system, auto-throttle, and a glideslope and approach track controller for longitudinal and lateral landing control,  respectively, and the environmental components including the atmospheric turbulence and carrier airwakes. The numerical implementation of the SCONE data in the simulation model given as a look-up table is also provided.  Section~\ref{simulation_results} presents numerical results for the two performance metrics on flight path control and reduced approach speed. Lastly, section~\ref{conclusions} summarizes the results and future work.

\section{Simulation Models}\label{simulation_models}
\begin{figure}
	\begin{center}
		\includegraphics[width=0.9\textwidth]{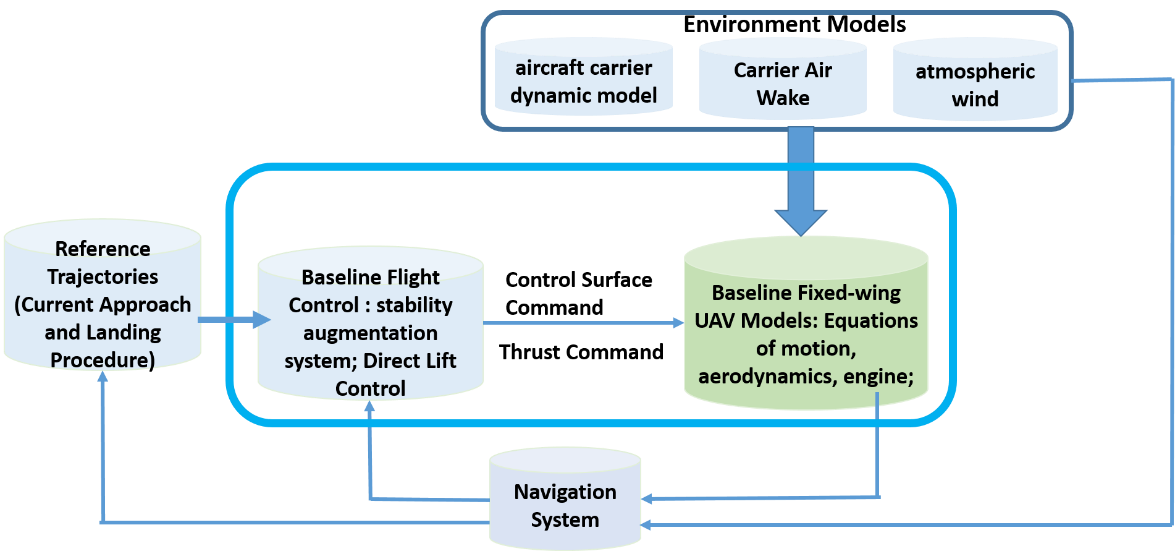}    
		\caption{System simulation models} 
		\label{sim_models}
	\end{center}
\end{figure}

The simulation models developed are schematically illustrated in Figure~\ref{sim_models}. The airborne components include baseline fixed-wing UAV models including equations of motion, aerodynamic models, engine models, and a baseline flight control system. The environment components include aircraft carrier dynamic model, atmospheric wind and carrier air wake. Current approach and landing procedures from~\cite{fitzgerald_flight_2004} are followed. Also because the objectives of this study are focused on reduced approach speed and landing performance, we only consider the segment after ‘tip over’ of the approach. The nominal glides slope will be set as a constant such as 3.5 degree. 
\subsection{Baseline Aircraft Models}
A model representative of F/A-18 E/F has been developed based on the F/A-18 High angle of attack (HARV) model~\cite{regenie1992f}. The physical parameters for the HARV model are shown in Table~\ref{table1}. 
\begin{table}[H]
\centering
\caption{Aircraft Parameters}
\label{table1}
\begin{tabular}{cccll}
Wing Area, S  &		400 ft$^2$  \\
Wing Span, b  &		37.42 ft  \\
Mean Aerodynamic Chord, c  &		11.52 ft  \\
Mass, m  &		1036 slug  \\
Maximum Thrust, $T_m$  &		11,200 lb  \\
Roll Moment of Inertia, $I_xx$  &		23,000 slug-ft$^2$  \\
Pitch Moment of Inertia, $I_yy$  &		151,293 slug-ft$^2$  \\
Yaw Moment of Inertia, $I_zz$  &		169,945 slug-ft$^2$  \\
\end{tabular}
\end{table}

The aerodynamic coefficients used in this study have been extracted from~\cite{fan1995time}. For carrier approach and landing configuration, the sea level altitude is considered for atmospheric properties . Assuming leading and trailing edge flaps completely down to 17.6 degrees and 45 degrees, respectively, and both left and right ailerons down to 42 deg, the aerodynamic coefficient dependencies are given as~\cite{fan1995time}.
\begin{align}
&C_{D} = \begin{cases}
\!\begin{aligned}
& 0.0013\alpha^{2}-0.00438\alpha  + 0.1423
\end{aligned}&  -5\leq \alpha \leq 20\\
\!\begin{aligned}
&-0.00000348\alpha^{2}+0.0473\alpha-0.3580\end{aligned} & \hspace{0.08in}20\leq \alpha\leq  40
\end{cases}\\
&C_{L} = \begin{cases}
\!\begin{aligned}
& 0.0751\alpha+0.0144\delta_{e}  + 0.732
\end{aligned}& \hspace{0.3in} -5\leq \alpha \leq 10\\
\!\begin{aligned}
&-0.00148\alpha^{2}+0.106\alpha+0.0144\delta_{e}+0.569\end{aligned} & \hspace{0.35in}10\leq \alpha\leq  40
\end{cases}\\
&C_{Y} = -0.0186\beta+\frac{\delta_{a}}{25}(-0.00227\alpha+0.039)+\frac{\delta_{r}}{30}(-0.00265\alpha+0.141)\\
&C_{m} = -0.00437\alpha-0.0196\delta_{e}-0.123q-0.1885\\
& C_{l} = C^{*}_{l}-0.0315p+0.0216r+\frac{\delta_{a}}{25}(0.00121\alpha-0.0628)-\frac{\delta_{r}}{30}(0.000351\alpha-0.0124)\\
&\text{where}\\
&C^{*}_{l} = \begin{cases}
\!\begin{aligned}
& (-0.00012\alpha-0.00092)\beta
\end{aligned}& \hspace{0.3in} -5\leq \alpha \leq 15\\
\!\begin{aligned}
&(0.00022\alpha-0.006)\beta\end{aligned} & \hspace{0.35in}15\leq \alpha\leq  40
\end{cases}\\
\end{align}
\begin{align}
&C_{n} = C^{*}_{n}-0.0142r+\frac{\delta_{a}}{25}(0.000213\alpha+0.00128)+\frac{\delta_{r}}{30} (0.000804\alpha-0.0474)\\
&\text{where}\\
&C^{*}_{n} = \begin{cases}
\!\begin{aligned}
&0.00125\beta
\end{aligned}& \hspace{0.3in} -5\leq \alpha \leq 10\\
\!\begin{aligned}
&(-0.00022\alpha+0.00342)\beta\end{aligned} & \hspace{0.35in}10\leq \alpha\leq  40\\
\!\begin{aligned}
&-0.00201\beta\end{aligned}& \hspace{0.35in}25\leq \alpha\leq  40
\end{cases}
\end{align}
where $\alpha$ and $\beta$ are the angle of attack and sideslip angle, respectively, $C_{D}, C_{L}, C_{Y}$ are the drag, lift and side force coefficients; $C_{l}, C_{m}$, and $C_{n}$ are the roll, pitch, and yaw-moment coefficients, respectively, and $\delta_{a}, \delta_{e}$ and $\delta_{r}$ are the aileron, elevator, and rudder deflections in deg. The simulation results shown in Section~\ref{simulation_results} consider aerodynamics till an angle of attack upto 40 deg. 

\subsection{Baseline Flight Control System }
\begin{figure}[H]
	\centering
		\includegraphics[width=0.85\textwidth]{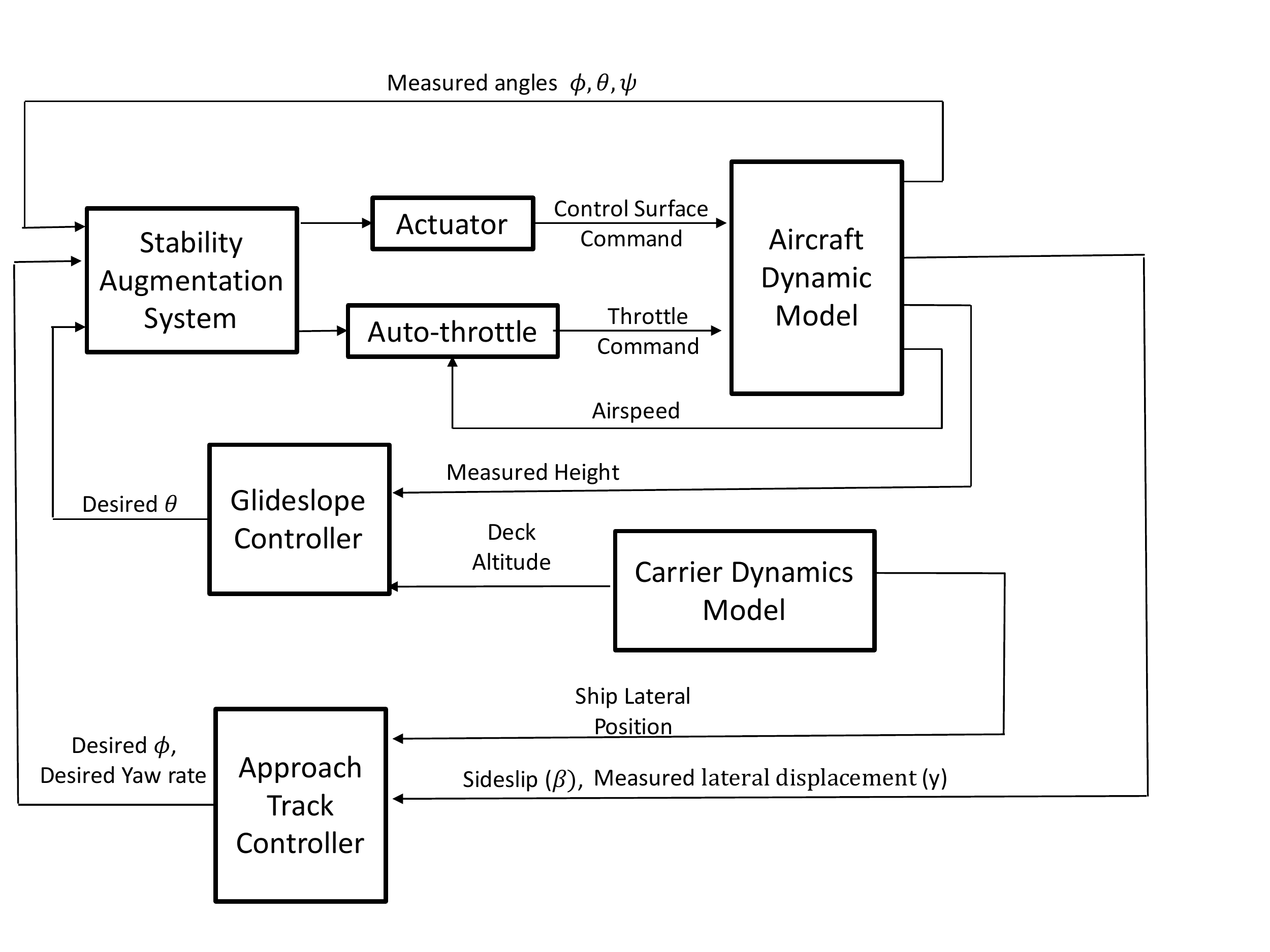}
	\caption{{Control system diagram}}
	\label{ControlSystem} 
\end{figure}

The baseline flight control system includes a glideslope controller, an approach track controller, a stability augmentation system (SAS) and auto-throttle. The diagram of the controls is shown in Figure \ref{ControlSystem}. The control system architecture proposed here is based on the flight control system presented in~\cite{regenie1992f}.
\subsubsection{Stability Augmentation System}
\begin{figure}[H]
	\centering
		\includegraphics[width=0.65\textwidth]{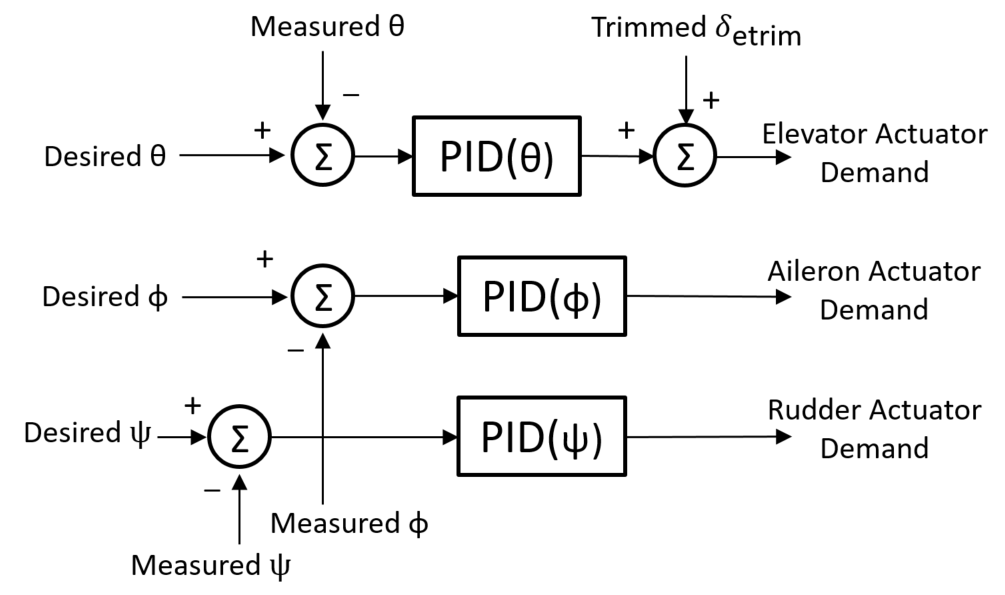}
	\caption{Stability augmentation system}
	\label{SAS} 
\end{figure}

Stability augmentation system(SAS) is commonly used to enhance the stability of an aircraft during flight. Proportional-derivative-integral (PID) controllers are implemented in the study to regulate the aircraft's attitude to the desired state. Given the desired aircraft attitude in terms of Euler angles $\phi_d, \theta_d, \psi_d$ and trimmed elevator deflection angle $\delta_{etrim}$ which is non-zero during the steady gliding, errors from desired and measured angles are the inputs of  the PID blocks. The structure of the SAS is shown in Figure \ref{SAS}.
The SAS gains are tuned using Simulink's control system design toolbox. The control laws of PID Controllers are listed below, where $\eta_{e_{ad}},\eta_{a_{ad}},\eta_{r_{ad}}$ are actuator demands of the elevator, aileron, and rudder, respectively and $ \epsilon_\theta$, $ \epsilon_\phi$, and $ \epsilon_\psi$ are the errors in pitch, roll, and yaw.
\begin{align}
\eta_{e_{ad}} =\delta_{etrim}+P_\theta \epsilon_\theta+I_\theta \int{ \epsilon_\theta\;dt}+D_\theta \frac{d}{dt}\epsilon_\theta
\end{align}
\begin{align}
\eta_{a_{ad}} =P_\phi \epsilon_\phi+I_\phi \int{ \epsilon_\phi\;dt}+D_\phi \frac{d}{dt}\epsilon_\phi
\end{align}
\begin{align}
\eta_{r_{ad}} =P_\psi \epsilon_\psi+I_\psi \int{ \epsilon_\psi\;dt}+D_\psi \frac{d}{dt}\epsilon_\psi
\end{align}
The tuned gains for the SAS used in the simulations in Section~\ref{simulation_results} are provided in the table below
\begin{table}[H]
	\centering
	\caption{Tuned SAS gains}
	\label{table2}
\begin{tabular}{cccll}
	$P_{\theta}$  &	-53.227	  \\
	$I_{\theta}$  &	-2.354  \\
	$D_{\theta}$  &	-97.452	 \\
    $P_{\phi}$  &	-14.997	 \\
    $I_{\phi}$  &	-0.7946	  \\
    $D_{\phi}$  &	-62.889	  \\
 $P_{\psi}$  &	179551.93	  \\
$I_{\psi}$  &	758023.01  \\
$D_{\psi}$  &3565.813	  \\
\end{tabular}
\end{table}
The SAS also includes second order actuator models for the elevator, aileron, and rudder and are expressed as~\cite{strickland_simulation_1996}
\begin{align}
\frac{\delta_{e}}{\delta_{c}} =&\frac{30.74^{2}}{s^2+2(0.509)(30.74)s+30.74^2}\\
\frac{\delta_{a}}{\delta_{c}} =&\frac{75^{2}}{s^2+2(0.59)(75)s+75^2}\\
\frac{\delta_{r}}{\delta_{c}} =&\frac{72.1^{2}}{s^2+2(0.69)(72.1)s+72.1^2}\\
\end{align}
where $\delta_{c}$ is the actuator input.
\subsubsection{Glideslope Controller}
\begin{figure}[H]
	\centering
		\includegraphics[width=0.5\textwidth]{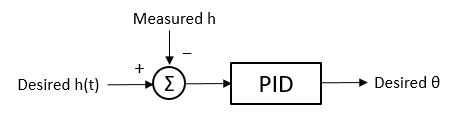}
	\caption{Glideslope controller}
	\label{VerticalController} 
\end{figure}

A glideslope based on the PID control law is implemented to stabilize the glideslope of the aircraft under disturbances. The control law is a function of the error in the aircraft's relative altitude to the carrier deck represented as $\epsilon_{h}$ in Eq.~\ref{gs}. Inspired by the approach glide path controller in \cite{regenie1992f}, we design the resulting control law based on the desired pitch angle ($\theta$) which is then regulated using the SAS described above. The structure of controller is shown in Figure \ref{VerticalController} and the corresponding gains are given in Table~\ref{glideslopegains}.
Additionally, from the obtained simulation results which will be explained later, it is observed that the error in the aircraft range (x) is typically small, under a low perturbation of wind. Therefore we do not explicitly consider a horizontal range controller in this paper, under the observation that the auto-throttle can regulate the airspeed in exponential time.
The gains for the vertical controller have been tuned in presence of atmospheric turbulence. The resulting control law is given as
\begin{align}\label{gs}
\theta_d =P_{gs}\epsilon_h+I_{gs}\int{ \epsilon_h\;dt}+D_{gs}\frac{d}{dt}\epsilon_h
\end{align}
\begin{table}[H]
	\centering
	\caption{Tuned approach track controller gains}
	\label{glideslopegains}
	\begin{tabular}{cccll}
		$ P_{gs}$  & -0.02736		  \\
		$I_{gs}$  &	-0.000959  \\
		$ D_{gs}$  &	-0.17342	 \\
		\end{tabular}
\end{table}
\subsubsection{Approach Track Controller}
The approach track controller design follows the methodology given in~\cite{fitzgerald_flight_2004}. The controller consists of two components: the aircraft's lateral position relative to the ship's lateral motion in a ground fixed reference frame is controlled via the ailerons through a desired roll command. In addition, a sideslip controller is implemented controlled via the rudder through a desired yaw rate command. The control laws are described as
\begin{align}
\phi_{d} = K_{py}y_{e}+	K_{iy}\int y_{e}dt+K_{dy}\frac{d y_{e}}{dt} \\
r_{d} = K_{p\beta}\beta_{e}+K_{i\beta}\int \beta_{e}dt+K_{d\beta}\frac{d\beta_{e}}{dt}
\end{align}
where $y_{e}$ is the error in the aircraft's lateral position relative to the ship, and $\beta_{e}$ is the sideslip error. The tuned gains are
\begin{table}[H]
	\centering
	\caption{Tuned approach track controller gains}
	\label{table3}
	\begin{tabular}{cccll}
		$ K_{py}$  & -0.02736		  \\
		$ K_{iy}$  &	-0.000959  \\
		$ K_{dy}$  &	-0.17342	 \\
		$K_{p\beta}$  &	9109.4	 \\
		$K_{i\beta}$  &	35962.11	  \\
		$K_{d\beta}$  &	308.67	  \\
		\end{tabular}
\end{table}
\subsubsection{Auto-throttle}
An auto-throttle control enables the regulation of the aircraft airspeed via the throttle. This is especially important during carrier landing operations where a fixed approach speed is usually required. The control law is obtained by enforcing the following first order error dynamics.
\begin{align}
\dot{V}+K_{u}(V-V_{T}) =0 
\end{align}
where $V$ is the measured airspeed, $K_{u}$ is the gain, and $V_{T}$ is the trim airspeed required to be maintained during landing. Substituting the equations for the derivative of the airspeed, the desired throttle response can be found as
\begin{align}
T_{d} = \frac{mK_{u}(V_{T}-V)+F_{drag}+mg\sin{\gamma}}{T_{max}\cos{\alpha}\cos{\beta}}
\end{align}
where $m$ is the aircraft mass, $F_{drag}$ is the drag force, $\gamma$ is the glideslope angle, and $T_{max} = 11200$ lb is the maximum thrust. The tuned gain $K_{u}$ used in the simulations is calculated as $73$.
\subsection{Environment Model}
The environment model described in Figure~\ref{sim_models} consists of carrier motion, atmospheric disturbance, and the carrier airwake model. The atmospheric disturbance model based on the guidelines in the Department of Defense handbook of flying qualities~\cite{DOD} includes three components: a turbulence mode, discrete wind gusts, and wind shear. The carrier airwake model consists of fours components: periodic ship-motion induced turbulence, steady carrier airwake, random free air turbulence, and random ship wake disturbance. 
\subsubsection{Turbulence Model}
\begin{figure}[H]
	\centering
\includegraphics[width=0.6\textwidth]{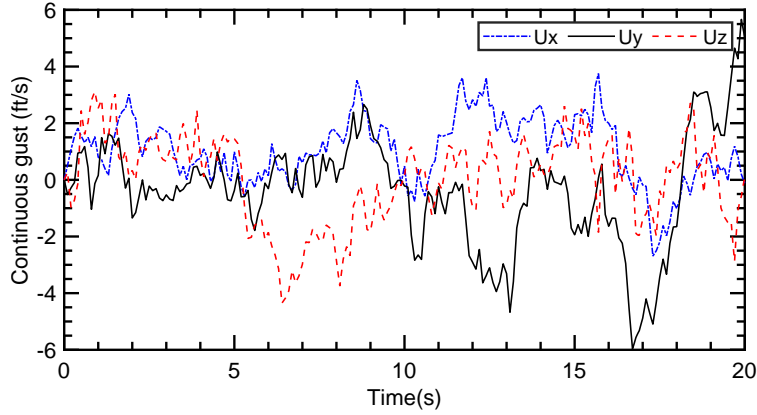}
\caption{Continuous gusts}
	\label{fig1} 
\end{figure}

 Continuous wind gusts considered here are spatially varying stochastic processes with Gaussian probability distribution. These gusts are commonly modeled using the Dryden or Von K\'{a}rm\'an models with their standardized forms provided in~\cite{DOD}. Both gust models are expressed in terms of its power spectral densities wherein the random processes are colored. 
The dryden form of the spectra for the gust components is given as
\begin{align}
\Phi_{u_{g}}(\Omega) = \sigma_{u}^{2}\frac{L_{u}}{\pi}\frac{1}{1+(L_{u}\Omega)^{2}}
\end{align}
\begin{align}
\Phi_{w_{g}}(\Omega) = \sigma_{w}^{2}\frac{L_{w}}{\pi}\frac{1+3(L_{w}\Omega)^2}{(1+(L_{w}\Omega)^{2})}
\end{align}
\begin{align}
\Phi_{q_{g}}(\Omega)= \frac{\Omega^{2}}{1+(\frac{4b\Omega}{\pi})^{2}}\Phi_{w_{g}}(\Omega)
\end{align}
where $\Phi_{u_{g}}$, $\Phi_{w_{g}}$, and $\Phi_{q_{g}}$ are the power spectral densities for translational ($u_{g}, w_{g}$) and rotary ($q_{g}$) gust components;  $\Omega$ is the spatial frequency, $\sigma_{u}$, $\sigma_{w}$ are the RMS turbulence intensities, $L_{u}$ and $L_{w}$ are the turbulence scales, and $b$ is the aircraft's wing span. For low altitude ($\sim$ 200 ft), the gust model parameters are taken as~\cite{DOD}
\begin{align}
L_{w} &= 100 \hspace{0.1in} \text{ ft} \\
L_{u} &= \frac{h}{(0.177+0.000823h)^{1.2}} \hspace{0.1in} \text{ ft}\\
\sigma_{w} &= 0.1W_{20} \hspace{0.1in} \text{ ft/s}\\
\sigma_{u}&=\frac{\sigma_{w}}{(0.177+0.000823h)^{0.4}} \hspace{0.1in} \text{ ft/s}
\end{align} 
where $h$ is the altitude in ft and $W_{20}$ is the wind speed at $20$ ft. 
Figure~\ref{fig1} shows the wind realizations at low turbulence.
\subsubsection{ Discrete Wind gusts}
\begin{figure}[H]
	\centering
	\includegraphics[width=0.6\textwidth]{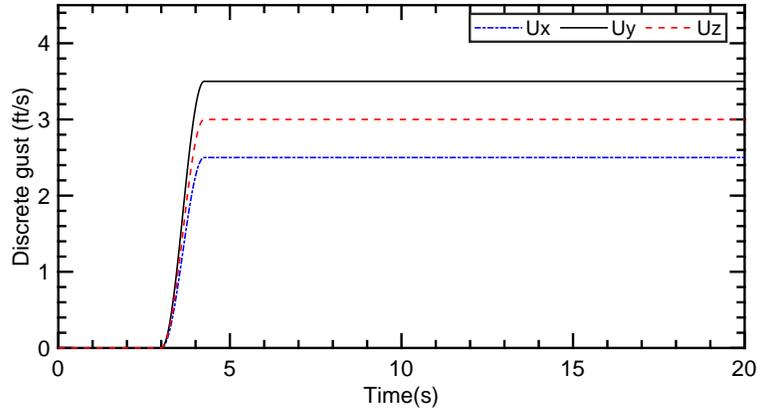}
	\caption{Discrete gusts}
	\label{discretegusts}
\end{figure}

The discrete gusts used in the study are based on a "1-cosine" model shown in Figure~\ref{discretegusts}. The inputs required to generate the gust are the gust lengths and gust magnitudes.
\begin{align}
&u_{x}, w_{x} =0, \hspace{0.2in} x<0\\
&u_{x}, w_{x}= \frac{V_{m}}{2}(1-\cos(\frac{\pi x}{d_{x,y,z}}))\hspace{0.2in} 0\leq x\leq d_{m}\\
& u_{x}, w_{x} =0 \hspace{0.2in} x>d_{m}
\end{align}  
where $u_{x}$ and $w_{x}$ are the discrete gust components in $x$ and $z$ axis of the aircraft body frame, $V_{m}$ is the gust amplitude, $d_{m}$ is the gust length defined as the point where the gust reaches it maximum, and $x$ is the distance traveled from the beginning of the simulation (starting from zero). The chosen parameter $V_{m}$ is $3.5$ ft/s for gust in x axis and $3$ ft/s for gust in z axis. The chosen gust length is $250$ in both x and z axis. The aircraft body frame is the coordinate fixed to the body, for which x-axis points to nose, y-axis points to right wing, and z-axis obeys to right-hand rule.
\subsubsection{Wind Shear}
The vertical wind shear in the vertical direction is calculated as
\begin{align}
W_{ws} = W_{20}\frac{\log(h/z_{0})}{\log(20/z_{0})}
\end{align}
where $W_{ws}$ is the vertical wind shear, $W_{20}$ is the wind velocity at 20 feet and is 15, 30, and 45 knots for low, moderate, and high turbulence, respectively. $z_{0}$ is specified as 0.15 ft for Category C flight phase which is the terminal phase. It should be noted that the $W_{20}$ is typically provided as a curve with respect to probability of occurrence with low and very low turbulence.The resultant mean wind shear in the inertial frame is changed to body-fixed axis coordinates.
\begin{figure}[H]
	\centering
	\includegraphics[width=0.6\textwidth]{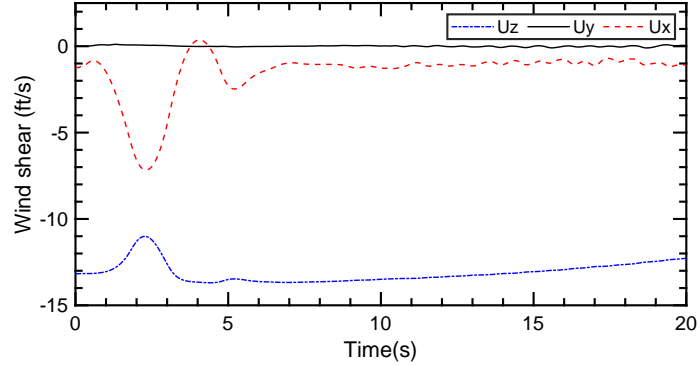}
	\caption{Vertical wind shear}
\end{figure}
\subsubsection{Periodic ship motion induced disturbance}
\begin{figure}[H]
	\centering
	\includegraphics[width=0.65\textwidth]{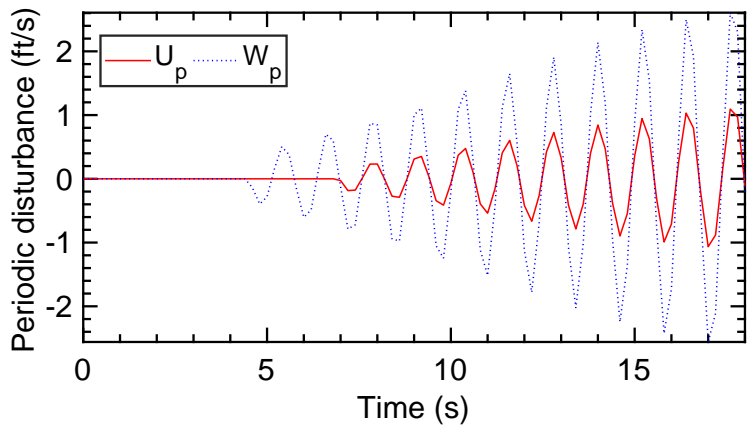}
	\caption{Periodic carrier turbulence}
	\label{periodic}
\end{figure}

The periodic wind disturbance which acts in the vertical and axial direction is a function of the velocity of wind over deck, the carrier's pitching frequency, pitch magnitude, and aircraft range. Wind over deck velocity is calculated as the difference between the nominal wind at sea and the (equal but opposite direction) of the ship velocity. 
\begin{align}
&U_{p} = \theta_{ac}V_{w/d}(2.22+0.0009X_{c})C\\
&W_{p} = \theta_{ac}V_{w/d}(4.98+0.0018X_{c})C\\
&C = \cos\bigg(\omega_{p}\big(t(1-\frac{V-V_{w/d}}{0.85V_{w/d}})+\frac{X_{c}}{0.85V_{w/d}}\big)+P\bigg)
\end{align}
where $U_{p}$ and $W_{p}$ are the axial and vertical wind disturbance, $X$ is relative aircraft position, $\theta_{ac}$ is ship pitch, $w_{p}$ is pitching frequency, $V$ is aircraft speed, $V_{w/d}$ is wind over deck, $P$ is a random phase. Note that periodic airwake for longitudinal direction is zero for range greater than 2236 feet and zero for vertical direction when range is greater than 2536 feet. Figure~\ref{periodic} illustrates the disturbance over a time span of $20$ s. 

For the simulation shown above, the range begins at 3142 feet, airspeed is taken as $225$ ft/s, ship speed is taken as 15 knots, nominal sea wind is taken as 5.4 knots, the pitch is assumed as 0.018 rad, pitching frequency is assumed as 0.62 rad/s, and random phase $P$ is taken as 0.25$\pi$.
\subsubsection{Steady carrier airwake}
\begin{figure}[H]
 	\centering
 	\includegraphics[width=0.6\textwidth]{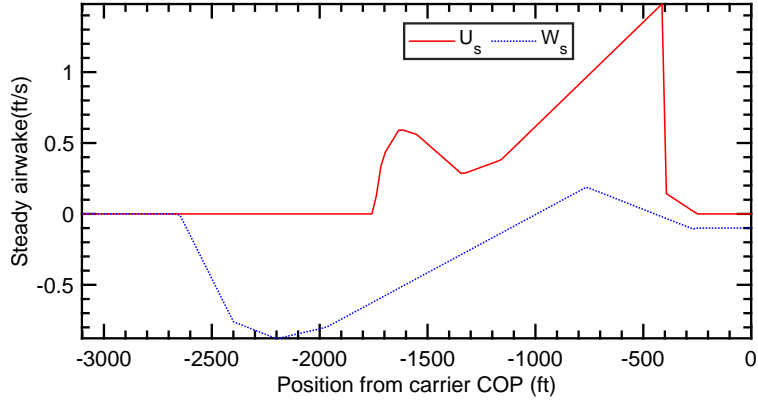}
 	\caption{Steady carrier airwake}
 	\label{steadyairwake}
 \end{figure}

 The steady component of the airwake is provided as a look-up table in terms of the ratio of the steady-wind ($\text{U}_{s}$ and $\text{W}_{s}$) over the wind over deck $\text{V}_{w/d}$ and the range from the carrier center of pitch (COP)\cite{fitzgerald_flight_2004} and is illustrated in Figure~\ref{steadyairwake}.
\subsubsection{Free air and random turbulence}
\begin{figure}[H]
	\centering
	\includegraphics[width=0.6\textwidth]{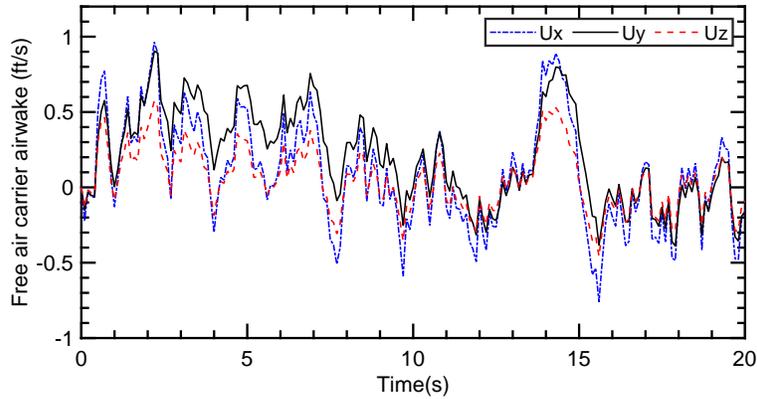}
	\caption{Free air turbulence}
	\label{freeair}
\end{figure}

The free air turbulence in the carrier disturbance model is calculated by passing white noise through a filter. This component is independent of the aircraft's relative position. The transfer functions to generate the wind are given as~\cite{DOD}
\begin{align}
\frac{u_{f}}{\eta} &= \sqrt{\frac{200}{V_{t}}}\frac{1}{1+\frac{100s}{V_{t}}}\\
\frac{v_{f}}{\eta} &= \sqrt{\frac{5900}{V_{t}}}\frac{1+\frac{400s}{V_{t}}}{(1+\frac{1000s}{V_{t}})(1+\frac{400s}{3V_{t}})}\\
\frac{w_{f}}{\eta} &= \sqrt{\frac{71.6}{V_{t}}}\frac{1}{1+\frac{100s}{V_{t}}}\\
\end{align}
where $V_{t}$ is the approach air speed, $u_{f}, v_{f}$ and $w_{f}$ are the turbulence in x,y and z axis, and $\eta$ is the band limited white noise. Figure~\ref{freeair} shows the variation of free air turbulence with time, where $V_t$ is taken as 218 ft/s for the simulation below. The random component of the airwake is also given as filtered white noise and is a function of the wind-over-deck, and the aircraft's relative position to the carrier. The random components are computed as
\begin{align}
\frac{U_{x}}{\hat{\eta}} &= \frac{\sigma(X_{c})\sqrt2\tau(X_{c})}{\tau(X_{c})s+1}\\
\frac{U_{z}}{\hat{\eta}} &= \frac{0.035V_{wd}\sqrt(6.66)}{3.33s+1}
\end{align}
where $\sigma (X_{c})$ and $\tau(X_{c})$ are provided as a look-up table. The white noise $\hat{\eta}$ is obtained as
\begin{align}
\hat{\eta} = \eta\frac{j\omega}{jw+0.1}\sin(10\pi t)
\end{align}

\begin{figure}[H]
	\centering
	\includegraphics[width=0.6\textwidth]{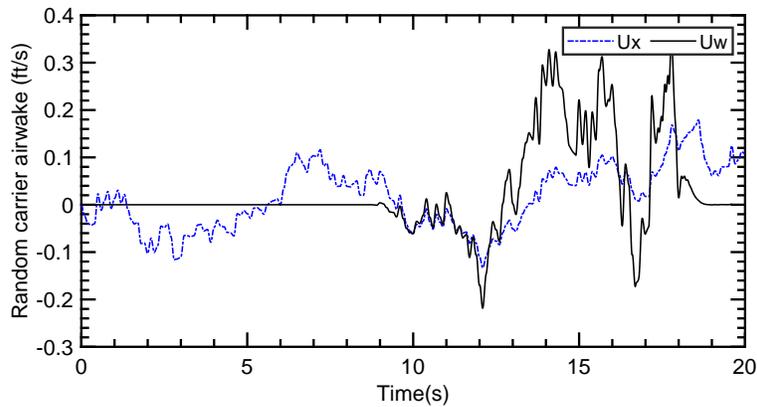}
	\caption{Random turbulence}
	\label{random}
\end{figure}
\subsection{Carrier Motion}
\begin{figure}[H]
	\centering
	\includegraphics[width=\textwidth]{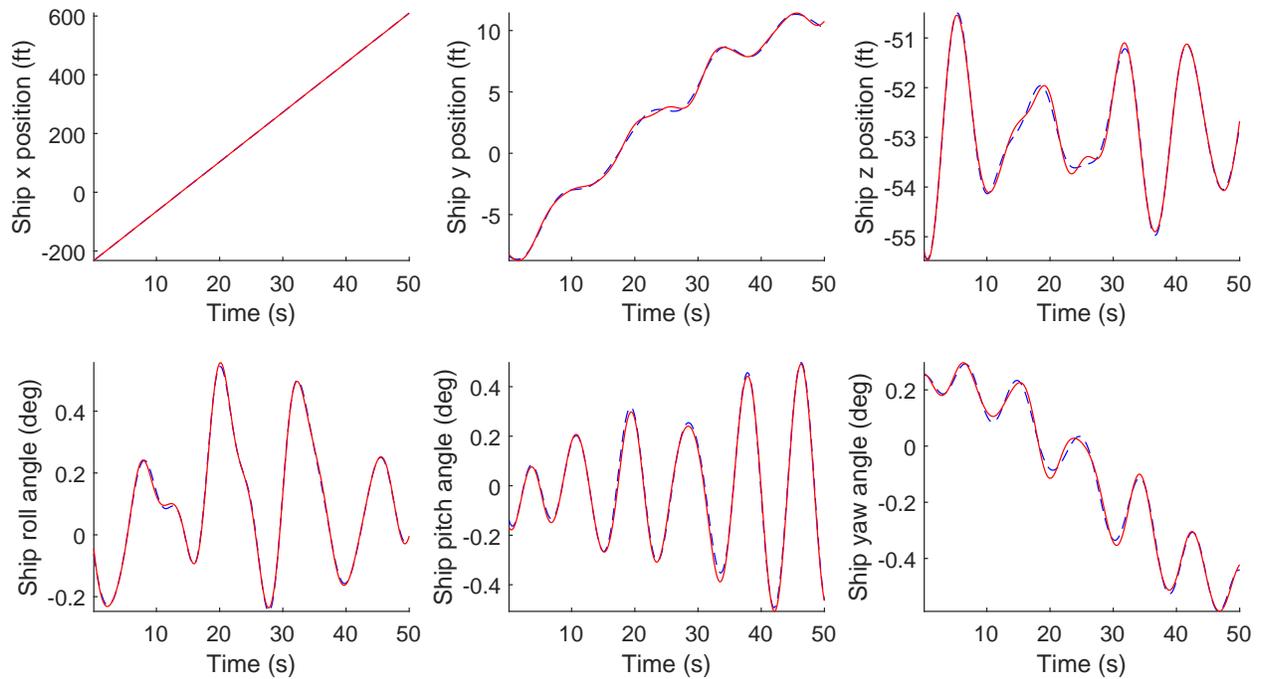}
	\caption{Nonlinear regression fit of Scone data with low sea state. Red and blue refer to the fitted and scone data, respectively.}
	\label{scone}
\end{figure}

\begin{figure}[H]
 	\centering
 	\includegraphics[width=0.6\textwidth]{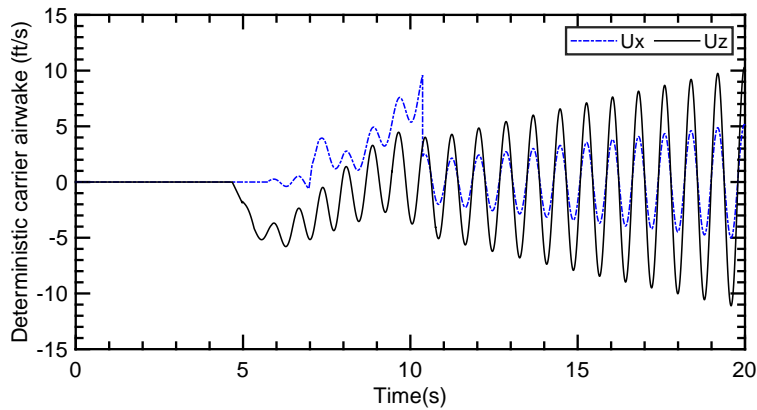}
 	\caption{Deterministic carrier airwake obtained using SCONE data}
 	\label{deterministic}
 \end{figure}

The carrier deck motion used in the simulations is based on the Systematic Characterization Of the Naval Environment (SCONE) data provided by the Office of Naval Research (ONR). The data includes
“low”, “medium”, and “high” deck motion cases for either roll or heave rate as the primary determinant of motion level. The data is provided as a look-up table for several sea-state levels over a total time frame of 30 minutes under a sampling rate of 20 Hz. The look-up table is not directly compatible with the Simulink based controller architecture developed since the propagation of the aircraft states uses a variable step size numerical ordinary differential equation (ODE) solver. Nonlinear regression tools are used to fit the available data into time-parametrized functions given mostly as either sum of sinusoids or Fourier series. Fig.~\ref{scone} shows the fitted profiles for the carrier's position and velocity for the heave dominated, low sea state.
 Using the numerical fit, the deterministic carrier airwake profile which includes both periodic and steady airwake is illustrated in Fig.~\ref{deterministic}.
\section{Simulation Results}\label{simulation_results}
Monte-carlo analysis is conducted to assess the controller performance in terms of the two performance metrics described below
\subsection{Flight path control performance}
\begin{figure}[H]
 	\centering
 	\includegraphics[width=0.85\textwidth]{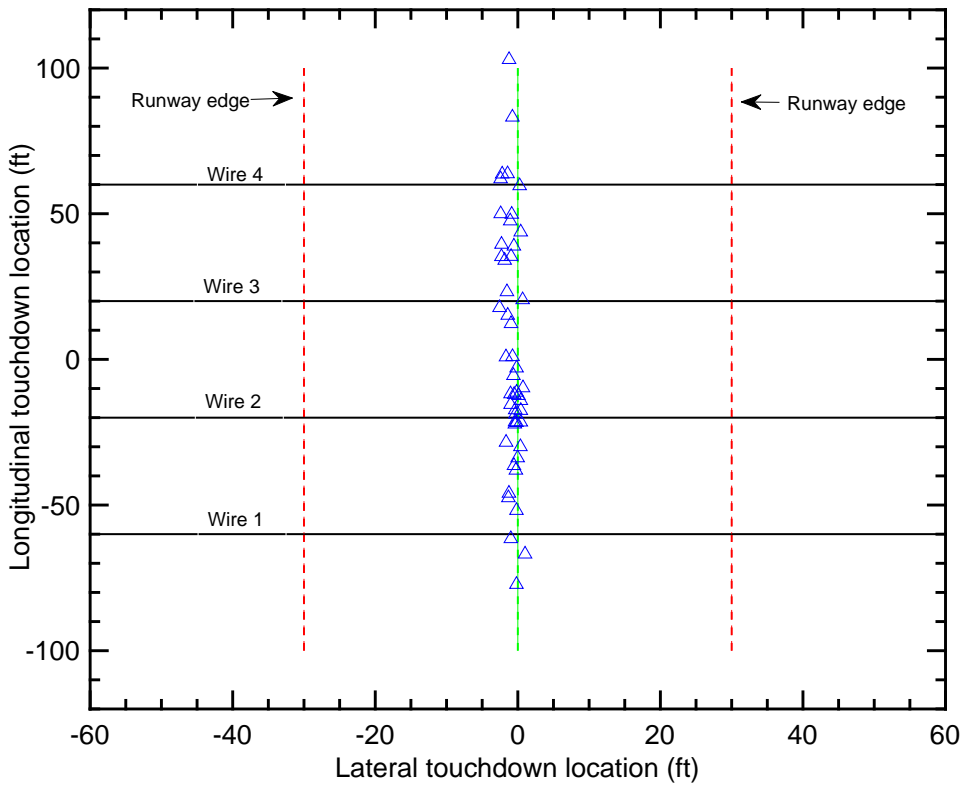}
 	\caption{Landing dispersion}
 	\label{landingdispersion}
 \end{figure}

\begin{figure}[H]
 	\centering
 	\includegraphics[width=\textwidth]{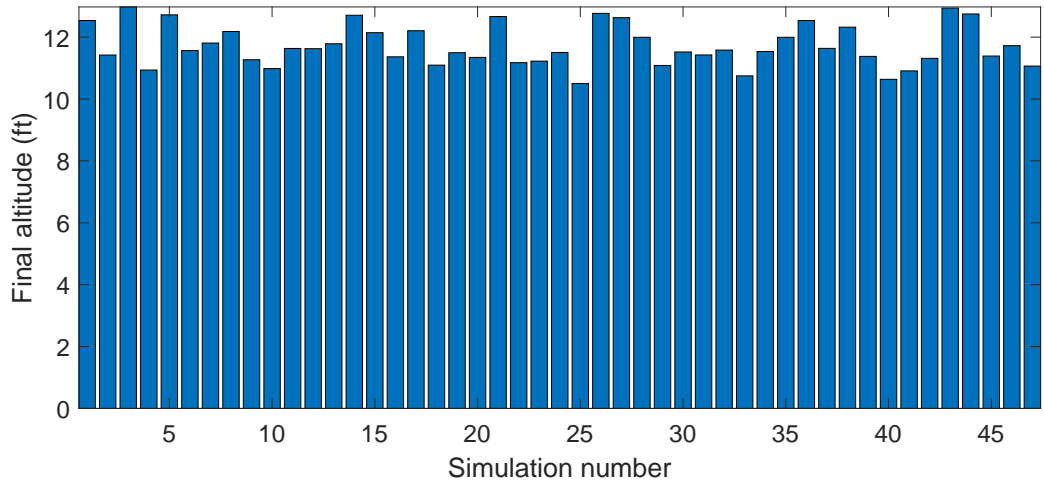}
 	\caption{Final altitude for successful landings}
 	\label{finalaltitude}
 \end{figure}

\begin{figure}[H]
	\centering
	\includegraphics[width=1.05\textwidth]{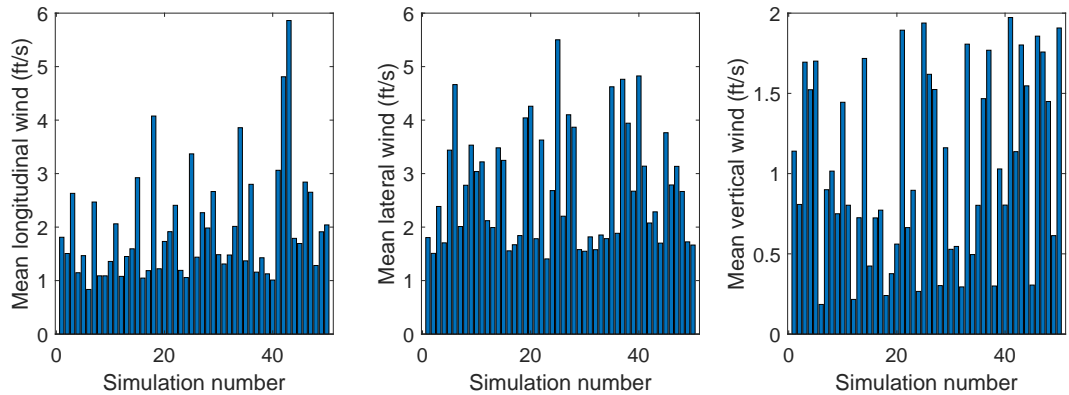}
	\caption{Mean wind speeds in x,y, and z directions}
	\label{meanwind}
\end{figure}

The flight control performance is analyzed by conducting 50 simulations with randomized initial conditions defined in Table~\ref{cond}. The wind environment consists of atmospheric turbulence including discrete, continuous gusts and wind shear, and carrier airwakes. The turbulence level considered is low. The sea-state considered is low and heave dominated as illustrated in Fig.~\ref{scone}. The noise seed taken for generating the stochastic component of gusts is sampled using uniformly distributed pseudo-random integers between 0-10$^5$. The control inputs were saturated before implementing them in the aircraft equations of motion. This is done to ensure that no control bound violations occur. As explained in the previous sub-section, nonlinear regression tools were used to provide time profiles of the ship motion from the SCONE data.
\begin{table}[H]
	\centering
	\caption{Simulation Setup}
	\label{cond}
	\begin{tabular}{cccll}
			Wind direction & Uniformly distributed pseudo-random integers between 0 to 180 degrees\\
		Continuous gust velocity & Uniformly distributed pseudo-random integers -8 to 8 ft/s\\
		Wind shear velocity & Uniformly distributed pseudo-random integers between -2 to 2 ft/s\\
		Discrete gusts &  Uniformly distributed pseudo-random integers between -2 to 2 ft/s\\
				Time of flight & 20 s\\
		Initial states & Trim, V$_{T}$ = 225 ft/s, $\alpha = 7.25 \deg$, $\theta = 3.81 \deg$. 	
		\end{tabular}
\end{table}
 Out of 50 simulations, 3 cases were reported to be failures, thereby making the landing success as 96\%. The deck landing dispersion is shown in Fig.~\ref{landingdispersion}. The landing area layout is based on a Nimitz class carrier and is obtained from~\cite{denison_automated_2007}.  
For most of the successful traps, the lateral dispersion is within $\pm$5 ft. On the contrary, the longitudinal dispersion is much larger. 2 bolter cases are reported, wherein the aircraft touches down on the deck but the hook misses all the wires. There are two cases where the aircraft touches down approximately 3 ft from the fourth wire. These are considered as successful landings since typically there is an error margin called the safe landing edge after the fourth wire. There is one case where the aircraft fails to clear the ramp in the given time and is not shown in the figure. The mean longitudinal landing position is -0.86 ft while the mean lateral landing position is 1.31 ft. The standard deviation for the longitudinal position is 37.2 ft while it is 0.92 ft for the lateral position. The final altitude is shown in Fig.~\ref{finalaltitude}. For all successful landings, the final altitude error is less than 13 ft. Fig.~\ref{meanwind} shows the average wind speed in x,y, and z directions. The mean wind magnitudes for most cases are reported to be below 6 ft/s which corresponds to low turbulence conditions.
\subsection{Reduced Approach Speed Analysis}
The carrier landing performance under the same baseline control laws is studied but with reduced approach speeds. To this end, eight trim conditions are numerically computed at different reduced approach speeds ranging from 150-200 ft/s. The trim states for the reduced approach speeds are given in Table~\ref{table5}.
\begin{table}[H]
	\centering
	\caption{Reduced approach speed trim conditions}
	\label{table5}
	\begin{tabular}{ccccccccccll}
		$V_{t}$ (ft/s) & 150 & 155 & 160 & 165 & 170 & 180&190&200&210&215&220 	  \\
		$\alpha$ (deg)& 23.3 &21.86&20.39&19.01&17.67&15.26&13.08&11.15&9.45&8.67&7.94\\
		$\theta$ (deg) &19.89&18.37&16.92&15.50&14.22&11.60&9.66&7.67&5.97&5.26&4.46 \\
		$\delta_{e}$ (deg) & -14.83&-14.49&-14.16&-13.85&-13.55&-13.02&-12.53&-12.01&-11.72&-11.55&-11.38 \\
		Thrust (lb) &8540&7900&7300&6720&6220&5200&4600&4000&3600&3500&3350
	\end{tabular}
\end{table}
The analysis is divided into three separate cases of the wind environment, namely, high fidelity wind model with atmospheric and carrier turbulence, atmospheric turbulence but without airwakes, and atmospheric turbulence and airwakes but without wind shear. The motivation behind choosing these three cases is to numerically assess the impact of individual wind components on the flight performance. For each wind scenario and eleven different approach speed, 50 simulations with random wind conditions corresponding to low turbulence shown in Table~\ref{cond}, are conducted, resulting in a total of 1650 Monte Carlo runs. For all the cases studied, the angle of attack is limited to $40$ $\deg$ due to the unavailability of aerodynamic data beyond this value. The landing success used in all the cases studied is based on the following criteria
\bi
\item The altitude error must be smaller than 15 ft.
\item The altitude error should be positive, as a negative error corresponds to a strike.
\item The final glideslope error must be less than 5 deg.
\item The final landing point must be on the deck and where the aircraft catches one of the four arresting wires. If the landing point misses the fourth wire by a margin greater than 5 ft, it is considered a bolter.
\item The final sink rate must be smaller than 12 ft/s. 
\ei 
\subsubsection{Case-I: Atmospheric turbulence and airwake}
\begin{figure}[H]
	\centering
	\includegraphics[width=\textwidth]{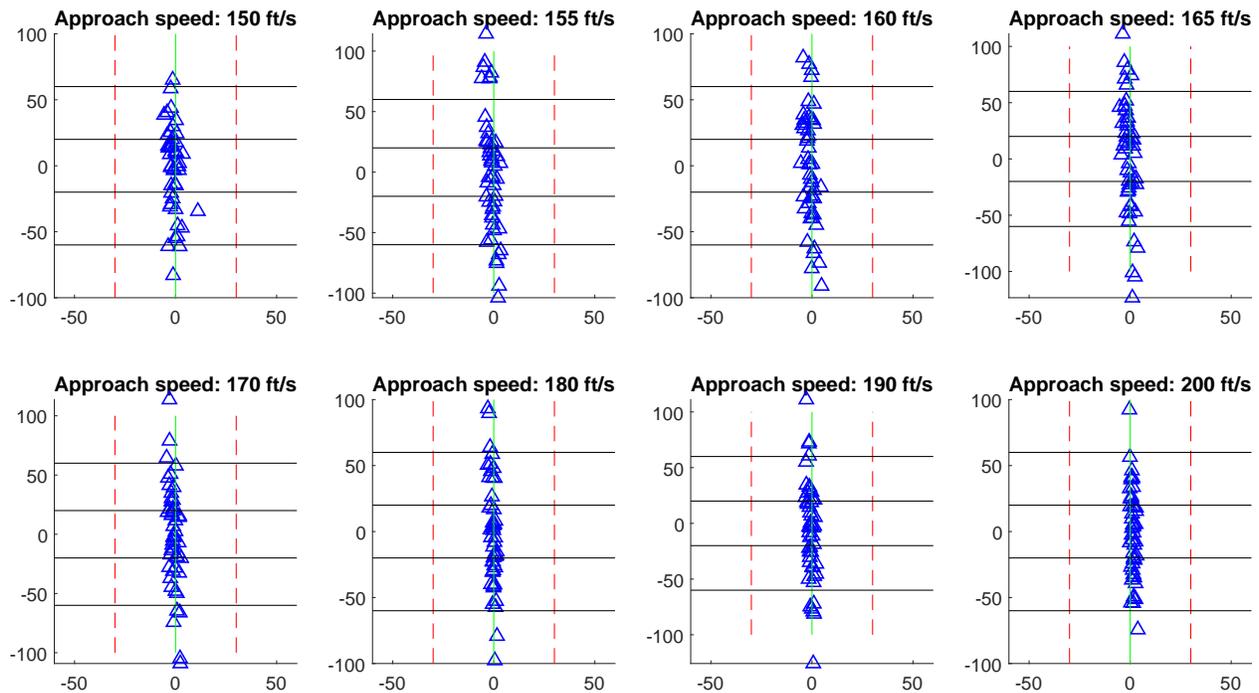}
	\caption{Landing dispersion}
	\label{landingdispersion_fullwind}
\end{figure}

\begin{figure}[H]
	\centering
	\includegraphics[width=\textwidth]{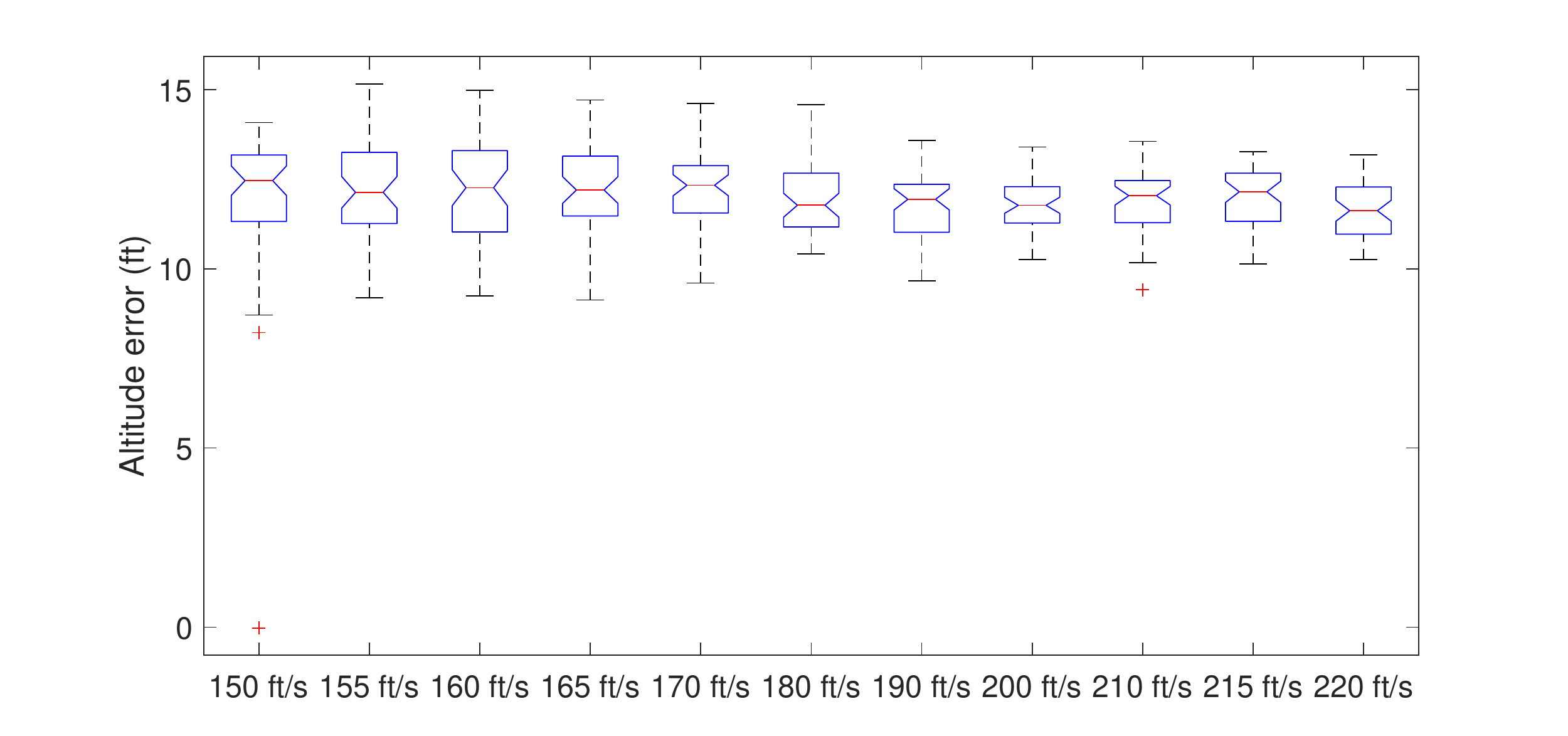}
	\caption{Altitude error}
	\label{alterror_fullwind}
\end{figure}

This case corresponds to a high fidelity model with continuous and discrete gusts, wind shear and carrier airwakes. A total of 50 simulations are conducted for each approach speed and the landing dispersion is recorded. Cases where the final landing point is outside the deck range are not shown in the Fig.~\ref{landingdispersion_fullwind}. In Table~\ref{table6}, we summarize the results in terms of success rate along with mean and standard deviation of final landing positions for successful wiretraps. In all subsequent result summaries, for the longitudinal position, the mean is denoted as $\mu_{x}$ and standard deviation as $\sigma_{x}$. For the lateral position, the mean is denoted as $\mu_{y}$ and standard deviation as $\sigma_{y}$. The final altitude error is shown in Fig.~\ref{alterror_fullwind}. The variance in the altitude error is observed to decrease with the approach speeds.
\begin{table}[H]
	\centering
	\caption{Result summary for reduced approach speeds}
	\label{table6}
	\begin{tabular}{ccccccccccll}
		$V_{t}$ (ft/s) & 150 & 155 & 160 & 165 & 170 & 180&190&200&210&215 &220\\
		Success rate&  88\%&84\%& 90\%& 86\%&92\% &92\%  & 90\%&98\%& 92\% &88\%&90\%  \\
		$\mu_{x}$ (ft) & -3.46&-12.58 & -8.37&-5&-7.77&-6.75&-13.1&-4.05&-9.55&-3.37&-9.59\\
		$\mu_{y}$ (ft) & 1.23& 1.37 & 1.23 &1.48&1.22&1.02&1.59&1.43&1.50&1.55&1.53\\
	  $\sigma_{x}$ (ft) &32.12 & 35.2& 36.67&40.32 &35.25 &35.73 & 34.74&31.4&39.16&42.75&40.2\\
			 $\sigma_{y}$ (ft) & 2.73 & 2.04&2.26&2.06&1.68&1.02&1.3&0.9&0.94&0.81&0.75 \\
		\end{tabular}
\end{table}
From the results, it is found that the success rate is typically high for all ranges of approach speed. The standard deviation in lateral dispersion is observed to decrease with increasing approach speeds. However,for the longitudinal dispersion, the standard deviation is not found to have a particular trend. In all the cases, the longitudinal landing dispersion is quite significant with the standard deviation $\sigma_{x}$ larger than 30 ft.

\subsubsection{Case-II: Continuous gusts and airwake}
\begin{figure}[H]
	\centering
	\includegraphics[width=\textwidth]{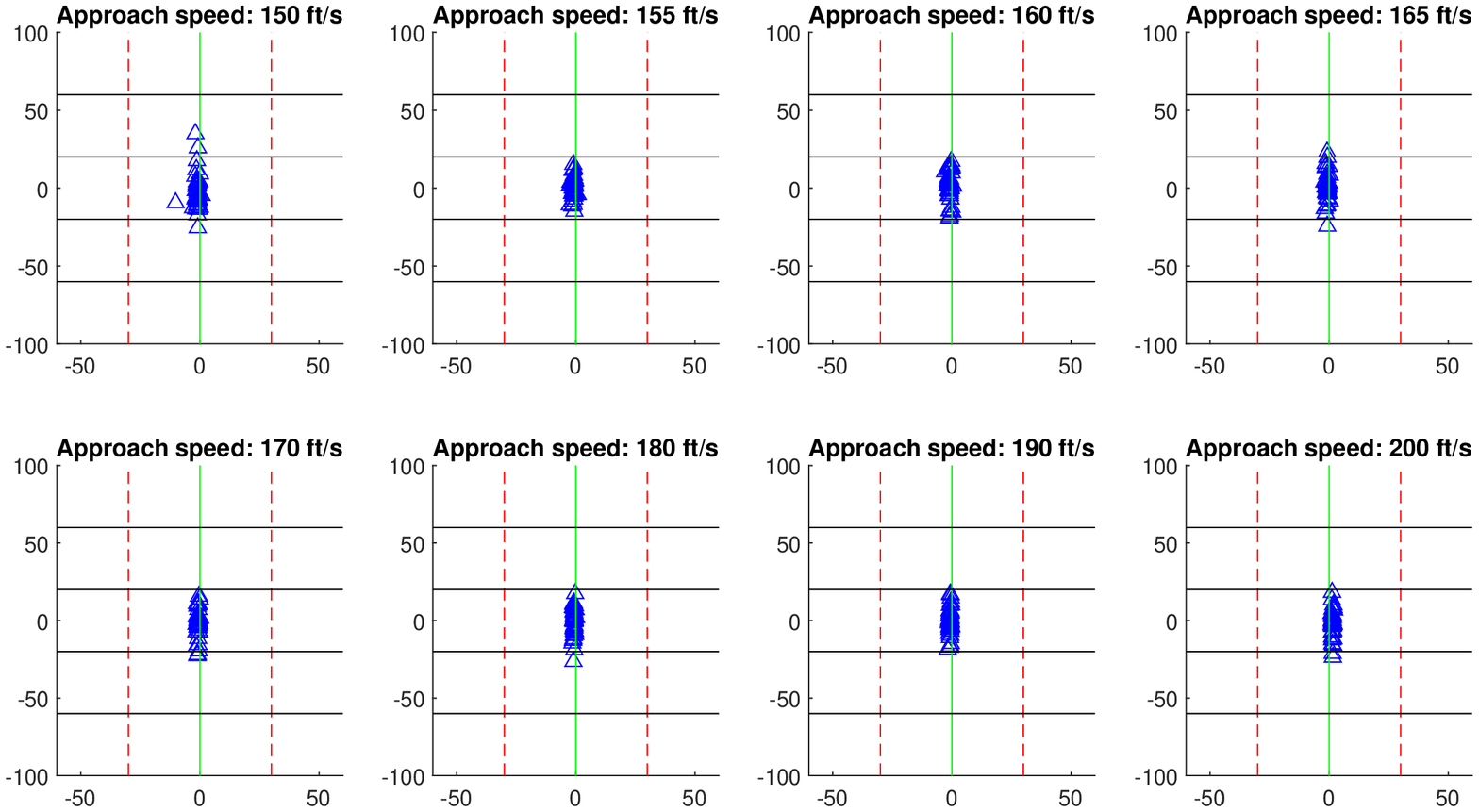}
	\caption{Landing dispersion}
	\label{landingdispersion_drydenburble}
\end{figure}

\begin{figure}[H]
	\centering
	\includegraphics[width=\textwidth]{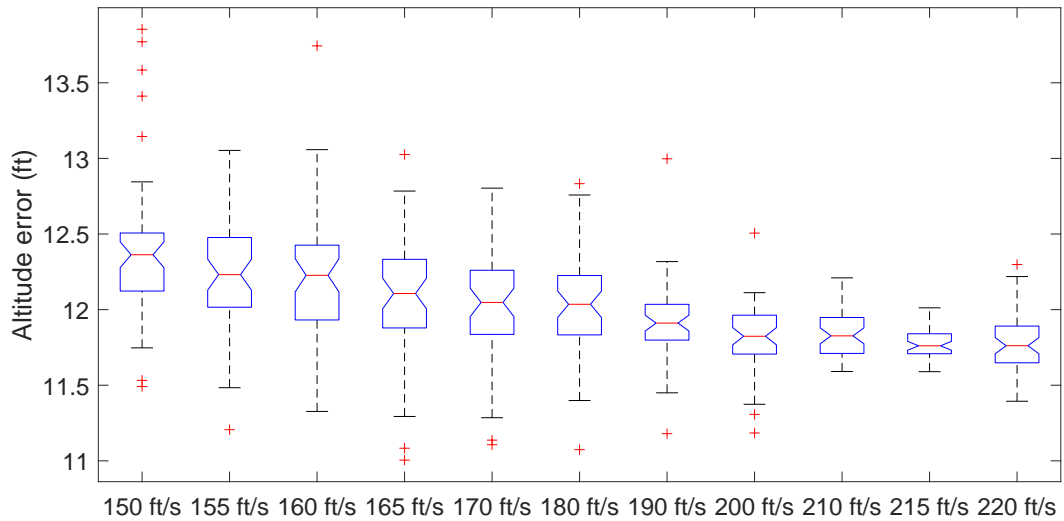}
	\caption{Altitude error}
	\label{alterror_drydenburble}
\end{figure}

In this case, the effect of wind shear and discrete gusts is not considered in simulations. Interestingly, the landing performance improves remarkably for all approach speeds with 100$\%$ success rate for all cases. This is due to the fact that both shear and discrete gust profiles are nearly constant throughout the flight and impact the airspeed more than continuous gusts and airwakes which are time-varying. For speeds 155-220 ft/s, all cases reported successful wire traps between the second and third arresting wire as shown in Fig.~\ref{landingdispersion_drydenburble}. The altitude error was also contained within 15 ft shown in Fig.~\ref{alterror_drydenburble}. As expected, the dispersion and altitude errors reduced as the approach speeds were increased. The standard deviation dispersion in both longitudinal and lateral plane are reduced with increased approach speeds as seen from Table~\ref{table7}. 
\begin{table}[H]
	\centering
	\caption{Success rate for reduced approach speeds}
	\label{table7}
	\begin{tabular}{ccccccccccll}
		$V_{t}$ (ft/s) & 150 & 155 & 160 & 165 & 170 & 180&190&200&210&215&220 	  \\
		Success rate & 100\% &100\% &100\%&100\%&100\%&100\%&100\% &100\%&100\%&100\% &100\% \\
	$\mu_{x}$ (ft) &-2.14 &0.93&2.03&0.87&-1.37&-0.34&-0.42&-1.94&-0.47&-0.67&-0.09 \\
$\mu_{y}$ (ft)  & 0.81&1.07 &1.1&1.16&0.32&1.27&2.28&1.46&1.42
&1.42&1.42\\
$\sigma_{x}$ (ft) &10.25 &6.36 &8.98&8.58&8.1&8.23&7.85&7.99&6.67&4.10&7.35\\
$\sigma_{y}$ (ft) & 1.4 &0.54 &0.57&0.45&0.32&0.26&0.25&0.21&0.14&0.09&0.19\\
	\end{tabular}
\end{table}
\subsubsection{Case-III: Wind shear, continuous, discrete gusts (No airwake)}
\begin{figure}[H]
	\centering
	\includegraphics[width=\textwidth]{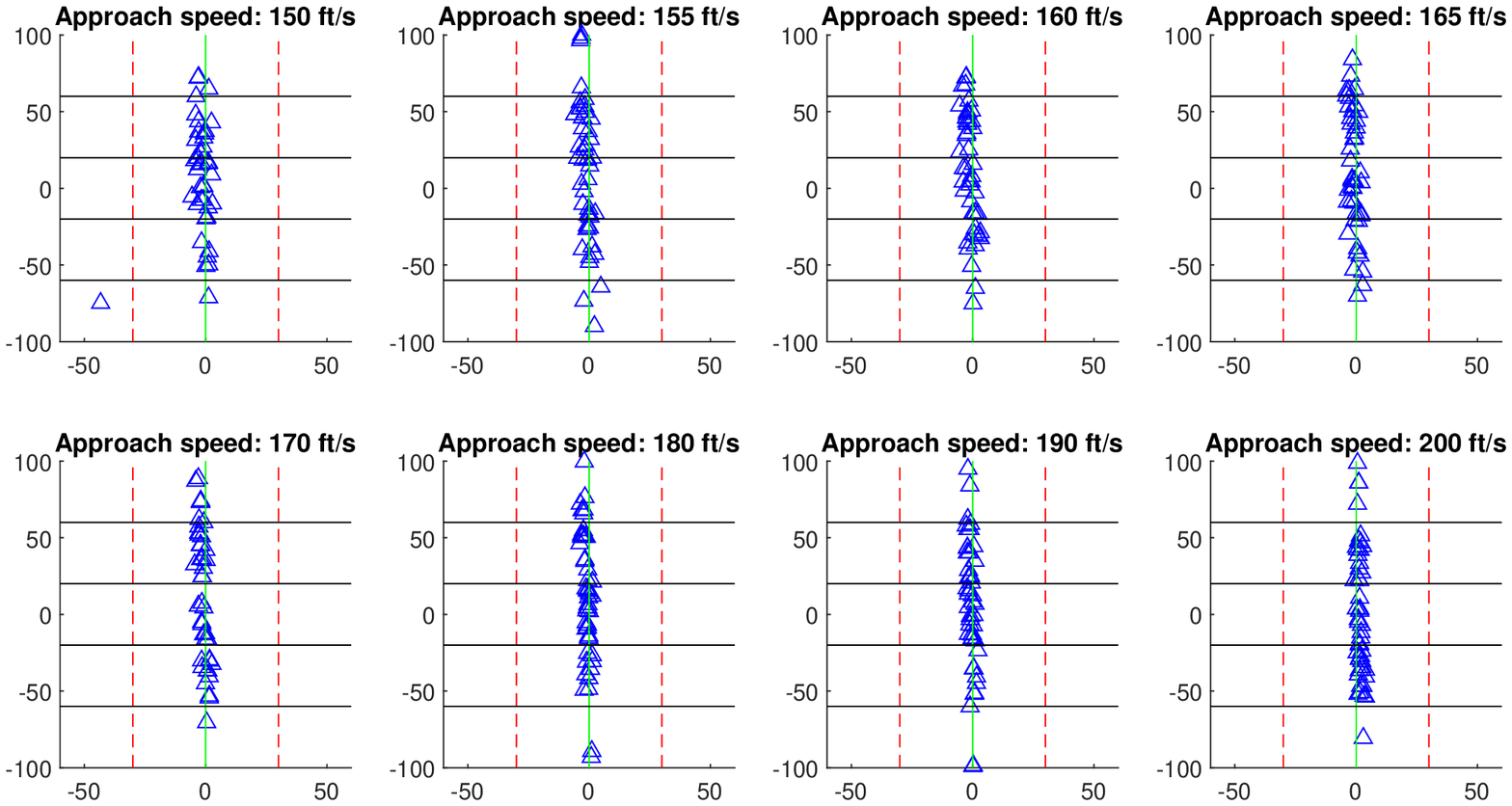}
	\caption{Landing dispersion}
	\label{landingdispersion_noburble}
\end{figure}

\begin{figure}[H]
	\centering
	\includegraphics[width=\textwidth]{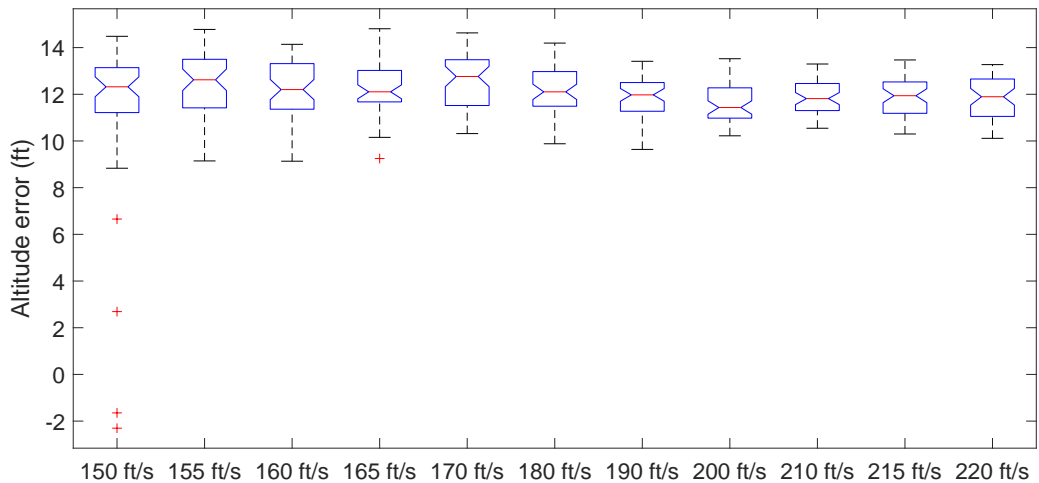}
	\caption{Altitude error}
	\label{alterror_noburble}
\end{figure}

This case corresponds to the calm sea state case where the effect of carrier airwakes is ignored. Typically, this would happen if the carrier is moving at very low speeds. The landing dispersion illustrated in Fig.~\ref{landingdispersion_noburble} shows worse performance compared to the both no wind shear and discrete gust case and the full-wind case. The error in the lateral direction is minimal with the mean less than 2 ft as seen from Table~\ref{table8} and most of the landing error is concentrated in the longitudinal direction. For the case of an approach speed of 150 ft/s, negative altitude errors are observed, as shown in Fig.~\ref{alterror_noburble}. This typically refers to the failure case of a rampstrike, wherein the aircraft strikes the ramp since the approach altitude was too low. 
\begin{table}[H]
	\centering
	\caption{Success rate for reduced approach speeds}
	\label{table8}
	\begin{tabular}{ccccccccccll}
		$V_{t}$ (ft/s) & 150 & 155 & 160 & 165 & 170 & 180&190&200&210&215&220	  \\
		Success rate&78\%& 86\%& 88\%  &84\% &86\%  &92\%& 92\%&92\% &88\%&92\%&88\%                              \\
		$\mu_{x}$ (ft) & 4.79&2.13&1.88&2.73&11.09&6.43&0.86&-6.02&0.04&-1.62&-2.39 \\
	$\mu_{y}$ (ft)  & 0.94&1.01&1.2&1.43&1.09&1.12&1.52&1.66&1.34&1.53&1.57\\
	$\sigma_{x}$ (ft) &30.3&42.12&37.87&34.45&40.92&39.18&44.29&36.84&32.14&41.07&44.41\\
	$\sigma_{y}$ (ft) &2.09&2.13&2.08&1.67&1.4&1.57&1.2&0.98&0.75&0.75&0.69\\		
	\end{tabular}
\end{table}
From the three cases analyzed, the impact of wind shear and discrete gust is found to be the highest on the landing performance. This intuitively makes sense since both shear and discrete gust profiles are typically constant throughout the time of flight considered. This effect is especially severe on the longitudinal dispersion while the altitude errors are still relatively contained. An approach speed of 150 ft/s is numerically found as the limiting case below which the landing success is limited. Any approach speed less than this limiting speed would typically have severe flight path performance degradation. It should be noted that from the aerodynamic analysis, the angle of attack hang-up condition where limited
pitch restoring moment is available to recover from a high angle of attack condition and can cause altitude loss is about 55 deg~\cite{wilt2003f}. In addition, the aerodynamic analysis in~\cite{chakraborty2010linear} shows the stall angle of attack, where the lift coefficient begins to decrease is approximately 40 deg. The corresponding approach speed for a 3.5 $\deg$ descent glideslope is~ 107 ft/s. The minimum approach we found is significantly larger than the 110\% stall based margins reported for minimum approach speeds and lower than the current recommended operational speed of approximately 225 ft/s. 
\section{Conclusion}\label{conclusions}
In this paper, a systematic control architecture is proposed and numerically validated for landings of fixed-wing unmanned aerial vehicles on aircraft carriers. The model considers a wind model consisting of atmospheric turbulence and carrier airwakes as well as carrier motion. Using Monte Carlo simulations, the performance of the flight path control system is studied using landing dispersion and altitude error results under low turbulence. In addition, this paper makes a case for reduced approach speed landings as well. Using smaller trim approach speeds under a range of wind conditions, 1650 simulations are conducted to numerically determine the limiting approach speed at which carrier landings can be performed. For the model considered based on the F/A-18 HARV aircraft, this limiting approach speed is larger than the reported stall margins and found to be 150 ft/s with a corresponding angle of attack of 21.8 $\deg$ and a -3.5 deg glideslope. 
\section*{Acknowledgments}
The authors acknowledge the research support from the Office of Naval Research (ONR) grant N00014-16-1-2729. The SCONE data, provided by ONR and the Naval Surface Warfare Center Carderock Division, are gratefully acknowledged.

\bibliography{carrier}

\end{document}